\documentclass[12pt]{article}

\usepackage{graphicx}
\usepackage{amsthm}
\usepackage{amsmath}
\usepackage{array}
\usepackage{natbib}
\usepackage{amsfonts}
\graphicspath{ {./images/} }
\usepackage{url} 

\newcommand{\blind}{0}

\begin{document}
\def\spacingset#1{\renewcommand{\baselinestretch}%
{#1}\small\normalsize} \spacingset{1}

\if0\blind
{
  \title{\bf Drift vs Shift: Decoupling Trends and Changepoint Analysis}
  \author{Haoxuan Wu,
    Department of Statistics and Data Science, Cornell University\\
    Toryn L. J. Schafer,
    Department of Statistics, Texas A\&M University\\
    Sean Ryan,
    and 
    David S. Matteson\thanks{
    The authors gratefully acknowledge \textit{The authors gratefully acknowledge financial support from the National Science Foundation 1455172, 1934985, 1940124, 1940276, 2114143, and USAID 7200AA18CA00014.}}\\
    Department of Statistics and Data Science, Cornell University}
  \maketitle
} \fi

\if1\blind
{
  \bigskip
  \bigskip
  \bigskip
  \begin{center}
    {\LARGE\bf Drift vs Shift: Decoupling Trends and Changepoint Analysis}
\end{center}
  \medskip
} \fi

\bigskip
\begin{abstract}
We introduce a new approach for decoupling trends (drift) and changepoints (shifts) in time series. Our locally adaptive model-based approach for robustly decoupling combines Bayesian trend filtering and machine learning based regularization. An over-parameterized Bayesian dynamic linear model (DLM) is first applied to characterize drift. Then a weighted penalized likelihood estimator is paired with the estimated DLM posterior distribution to identify shifts. We show how Bayesian DLMs specified with so-called shrinkage priors can provide smooth estimates of underlying trends in the presence of complex noise components. However, their inability to shrink exactly to zero inhibits direct changepoint detection. In contrast, penalized likelihood methods are highly effective in locating changepoints. However, they require data with simple patterns in both signal and noise. The proposed decoupling approach combines the strengths of both, i.e.\ the flexibility of Bayesian DLMs with the hard thresholding property of penalized likelihood estimators, to provide changepoint analysis in complex, modern settings. The proposed framework is outlier robust and can identify a variety of changes, including in mean and slope. It is also easily extended for analysis of parameter shifts in time-varying parameter models like dynamic regressions. We illustrate the flexibility and contrast the performance and robustness of our approach with several alternative methods across a wide range of simulations and application examples.
\end{abstract}

\noindent%
\vfill

\newpage
\spacingset{2}

\section{Introduction} \label{sec_intro}
Complex non-stationary dynamic systems often exhibit both global (macro patterns) and local (micro fluctuations) features of inferential interest. Herein, we focus on making distinctions between drift and shifts. Drift describes the micro-level evolution of a process and may appear as variation about gradual trends. In contrast, shifts represent macro-level changes in a process perceived as sharp discontinuities, rapid changes, or major breaks.

A commonly used approach for modeling drift in time series regression is the dynamic linear model (DLM). DLMs tend to be overparameterised models with at least one parameter per observation time. Therefore, we focus on Bayesian estimation of DLMs \citep{chan2018bayesian} with priors for selection or regularization. Continuous shrinkage priors regularize parameters to produce smoother and more reliable estimates for the dynamic features by ``shrinking" small values closer to zero \citep{Carvalho_2009, bitto2019achieving}. Bayesian DLMs with shrinkage priors are excellent for capturing changes that occur smoothly over time. However, since the inference does not include exact zero values, characterizing shifts in the trend, such as changepoints, is not straightforward. 

On the other hand, changepoint methods tend to be effective at capturing sudden breaks \citep{aminikhanghahi2017survey}. Common changepoint methods include likelihood ratio tests with cumulative statistics \citep{jeske2019cusum, Fryzlewicz_2014}, penalized likelihood approaches  \citep{Killick_2012, maidstone2017optimal} and non-parametric distanced based metrics \citep{Matteson_2013, james2014ecp}. While these methods have shown to be effective on well-behaved time series, they tend to struggle when we model systems characterized by drift and shift. 

Given the complementary strengths of Bayesian DLMs and changepoint models, it is natural to explore an intersection of these methods. In this paper, we propose a two step Bayesian method using a decoupled posterior summary that allows us to identify changepoints in any Bayesian DLM. First, a Bayesian DLM is fitted to filter the signal of the data from the noise components. We do not specify a particular structure for the DLM but rather will show the approach works for a wide range of structures. Second, a penalized loss on the posterior of the model imposes a sparse summary of changepoints locations. 

The decoupled approach as presented by \cite{Hahn_2015} separated the processes of regression modeling and discrete inference of variable selection. In a similar vein, \citet{huber2021inducing} applied the framework to a time-varying parameter model with a specification of the decoupled loss as introduced by \citet{ray2018signal}. In this paper, we extend the decoupled approach to non-stationary time series analysis and changepoint detection. 

The decoupled approach provides two key advantages. First, the decoupled approach separates the estimation of the trend from the changepoint locations. As a result, we can fit a highly flexible Bayesian model to deal with the intricacies of the data such as outliers, heterogeneity and seasonality. Most existing changepoint algorithms struggle to deal with these components as they tend to significantly skew the distribution of the data and violate distributional assumptions. Second, by using a penalized loss on the posterior, the decoupled approach is able to provide uncertainty estimates for the number of changepoints selected. In turn, the decoupled approach can provide more insights into the selection process and the trade-off between goodness-of-fit and the number of changepoints. 

The paper proceeds as follows. In Section \ref{sec_methd}, we introduce the decoupled approach for identifying changepoints in Bayesian DLMs. Section \ref{sec_sim_dec} and \ref{sec_real_w} illustrates the effectiveness of the decoupled approach in diverse sets of simulation scenarios and real-world datasets. We conclude with a discussion of key benefits. The Appendix details the loss derivation, methodology extensions, extended simulation results, and more real data applications.

\section{Methodology} \label{sec_methd}
\subsection{Decoupled Modeling} \label{sec_rw}

To introduce the decoupled approach, we start by introducing a standard Bayesian dynamic linear model (DLM). Suppose we observe a univariate time series $\pmb{Y} = (y_1, \dots, y_n)'$ and a predictor series $\pmb{X} = (x_1, \dots, x_n)'$, a Bayesian DLM can be formulated as follows: 
\begin{equation}\label{eq1}
    \begin{gathered}
        y_{t} = x_t \beta_t + \epsilon_t, \quad \epsilon_t \sim N(0, \sigma_{\epsilon, t}^2),\\
        \bigtriangleup^D \beta_{t} = \omega_{t}, \quad \omega_t \sim N(0, \sigma_{\omega}^2), 
    \end{gathered}
\end{equation}
where $\bigtriangleup^D (\cdot)$ is the degree $D$ differencing operator with $D = 0$ defined as the identity function. In this setup, $\{\beta_t\}$ encodes the time-varying relationship between the predictor series $\{x_t\}$ and the response series $\{y_t\}$. The process $\{\epsilon_t\}$ models noise; $\{\sigma_{\epsilon, t}^2\}$ is modeled as potentially time varying; a heteroskedastic noise process gives additional flexibility with low computational cost, in practice. For now, we will assume only one predictor series. Later on, we will extend the framework to deal with multiple predictors. 

Specifically, the random walk process, corresponding to $D=1$, induces smooth estimates for $\{\beta_t\}$ when $\sigma^2_{\omega}$ is small. For well-behaved time series, a globally smooth estimate for $\{\beta_t\}$ provides sufficient inference. However, \eqref{eq1} does not include a mechanism for discrete inference applicable to time series characterized by shifts. Rather than adjust the priors, we chose to take a decoupled approach to summarize the posterior. The decoupled approach summarizes a relatively smooth estimate of $\{\beta_t\}$ with a penalized loss function that induces discrete inference. The discrete inference explored by \citet{Hahn_2015} was variable selection and we adapt the approach for discrete shift features such as abrupt changepoints. As we will show later on, for more noisy series, locally adaptive shrinkage priors may be necessary to induce sufficiently smooth estimation in time series with variable degrees of wiggliness.

To illustrate the connection between variable selection and changepoint detection, notice that the time-varying relationship $\{\beta_t\}$ can be seen as a discrete integration over the estimated increments $\{\omega_t\}$ and the initial values of $\{\beta_t\}$. In order for the coefficient function to be constant for some period of time, the increments must be zero. Therefore, shift detection is equivalent to estimating the non-zero increments analogous to estimating the non-zero coefficients in variable selection inference. For the decoupled approach, we fit the above model to the observed data via Gibbs sampling with the MCMC sampling scheme provided by the R package \textit{dsp} from the methods in \citet{kowal_2018} to estimate the posteriors for the coefficients which are dense and non-zero everywhere by model construction. Then, we choose a penalized loss function to summarize the posterior. Due to heteroskedastic noise in \eqref{eq1}, we consider a weighted least squares loss function: $ L^*_\lambda(\widetilde{\pmb{y}}, \widetilde{\pmb{\beta}}) = \sum_{t=1}^n [w_t (\tilde{y}_t - x_t \widetilde{\beta}_t)]^2 + q_\lambda(\widetilde{\pmb{\beta}}),$ where $\widetilde{\pmb{y}}= (\widetilde{y}_1, \dots, \widetilde{y}_n)$ is the posterior prediction given \eqref{eq1}, $\widetilde{\pmb{\beta}} = (\widetilde{\beta}_1, \dots, \widetilde{\beta}_n)$ is the penalized linear predictor, $q_\lambda(\cdot)$ is a penalty function to induce sparsity given a penalty parameter $\lambda$, and $\{w_t\}$ are the weights for each time-step. Details on $q_\lambda()$ and $\{w_t\}$ will be given in Section \ref{subsec_wpf}. 

As in \citet{Hahn_2015}, we first integrate $L^*_\lambda(\widetilde{\pmb{y}}, \widetilde{\pmb{\beta}})$ over $\{\tilde{y}_t\}$ given $\{\beta_t, \sigma_{\epsilon, t}^2\}$ then integrate over $\{\beta_t, \sigma_{\epsilon, t}^2\}$ given $\{y_t\}$. This results in the decoupled loss as follows:
\begin{equation}\label{dec_loss}
    L^*_\lambda(\widetilde{\pmb{\beta}}_t) = \sum_{t=1}^n w_t (x_t \bar{\beta}_t - x_t \widetilde{\beta}_t)^2 + q_\lambda(\widetilde{\pmb{\beta}}_t),
\end{equation}
where $\{\bar{\beta}_t\}$ denotes the posterior mean of the trend estimate from the Bayesian DLM (Appendix 1). Equation \eqref{dec_loss} can be thought of as a second level shrinkage on the underlying coefficients to induce hard thresholding \citep{Hahn_2015}. The loss function, parameterized by the penalty  parameter $\lambda$, will be utilized to select changepoints from the posterior estimates of a Bayesian DLM. 

\subsection{Weights and Penalty Function} \label{subsec_wpf}

The choice of weighted least squares allows the approach to utilize the estimated variance from the Bayesian DLM to induce additional localized adaptivity. The weights adjust the penalty to time-varying volatility inherent in the data, inducing a smaller loss for time-steps with a larger variance and a larger loss for time-steps with a smaller variance. %This makes sense as we expect the process $\{\beta_t\}$ to have more variability in regions of high volatility. 

For the weights $\{w_t\}$, the classic choice is inverse to the noise \citep{kiers1997weighted}. In our case, since we have posterior estimates of the variance after sampling the Bayesian DLM, we set our weights to be 
\begin{equation*} \label{eq_weight}
w_t = \overline{\sigma_{\epsilon, t}^{-2}}, \quad \text{for} \ t = 1, \dots, n,
\end{equation*}
where $\overline{\sigma_{\epsilon, t}^{-2}}$ is the posterior mean for the precision at time $t$. %which will be utilized as our point estimate for the variance at time $t$. 
As previously discussed, the weights induce additional robustness for change detection in heteroskedastic data.

The penalty term $q_\lambda(\widetilde{\pmb{\beta}})$ will  penalize the number of time-steps for which the $D$th difference (i.e. D = 1 or 2) in $\widetilde{\pmb{\beta}}$ is non-zero. Table \ref{tab_pen} shows three possible choices for the penalty function. Ideally, the $\ell_0$ penalty will be used to identify the optimal subset of time-steps in which the $D$th difference are non-zero. However, the $\ell_0$ penalty is difficult to estimate efficiently, making it infeasible for long time-series. One solution is to relax the $\ell_0$ penalty to the $\ell_1$ penalty. However, the $\ell_1$ penalty induces shrinkage which tends to bias the result. This is due to the fact that the penalty term increases linearly in relation to $\{\bigtriangleup^D \widetilde{\beta}_t\}$, resulting in the penalty favoring changepoints of low magnitude. As we will show in the simulations, using the $\ell_1$ penalty directly will lead to significant over-estimation of changepoints.

\begin{table*}
\setlength{\baselineskip}{1.0pt} 
\caption{Choices of Penalty Function}
\label{tab_pen}
\normalsize
\centering
\begin{tabular}{|c | c c c|}
\hline 
   & $\ell_0$ & $\ell_1$ & Adaptive $\ell_1$  \\ 
 \hline 
 Penalty Function & $\lambda \sum_t \mathbb{I}_{\bigtriangleup^D \widetilde{\beta}_t \neq 0}$ & $\lambda \sum_t |\bigtriangleup^D \widetilde{\beta}_t|$ & $\lambda \sum_t \frac{1}{|\psi_t|} |\bigtriangleup^D \widetilde{\beta}_t|$ \\
  Motivation & Selection & Shrinkage and Selection & Combination of $\ell_0$ and $\ell_1$  \\
 \hline 
\end{tabular}
\begin{flushleft}
  \setlength{\baselineskip}{1.0pt} 
  {Table defines various penalty functions, $q_\lambda(\cdot)$, for differenced coefficient vectors and they're associated motivation. For example, the $\ell_0$ penalty can be used for selecting non-zero increments of the differenced coefficient vectors.}
\end{flushleft}
\end{table*}

A refined goal is then to identify a penalty function that combines the computational efficiency of the $\ell_1$ penalty and the optional subset selection ability of the $\ell_0$ penalty. As a result, we propose a version of the  adaptive $\ell_1$ penalty \citep{Zou_2006} that pushes the $\ell_1$ penalty closer to the $\ell_0$ penalty. The resulting penalty can be written as follows:
\begin{equation}\label{pen}
q_\lambda(\widetilde{\pmb{\beta}}) = \lambda \sum_t \frac{1}{|\psi_t|} |\bigtriangleup^D \widetilde{\beta}_t|,
\end{equation}
where $\psi_t = \overline{\bigtriangleup^D \beta_t}$ for all $t$. Motivated by similar refinement in \citet{Hahn_2015}, $\psi_t$ at time $t$ is the posterior mean of the $D$th degree difference of $\beta_t$. This term can function as a normalizer which levels the impact of each changepoint regardless of the magnitude of the change at that time-step. In a time-step with a larger change in $\{\beta_t\}$, $\psi_t$ will tend to be higher in magnitude, leading a changepoint to be penalized less. As a result, this weight term corrects some of the bias in the $\ell_1$ penalty and gives better results for changepoint estimation. Optimization of \eqref{pen} simplifies to using a standard penalized regression function; we used \textit{glmnet} in R \citep{tay2023elastic}. A computational special case of optimization without a global penalty, $\lambda$, is dicussed in \citep{ray2018signal}.

\subsection{Selecting the Optimal Number of Changepoints} \label{sec_diag_tool}
As seen in \eqref{dec_loss} and \eqref{pen}, we utilize a penalty function indexed by a parameter $\lambda$. The value of $\lambda$ plays a critical role in the final selection of the number of changepoints. As $\lambda$ approaches 0, there would be no enforcement of sparsity and every point will be treated as a changepoint. As $\lambda$ approaches $\infty$, all $\{\bigtriangleup^D \widetilde{\beta}_t\}$ will be 0 and no changepoint will be detected. Typically, with a penalized loss function, cross-validation is used to select the penalty parameter. However, in the case for the proposed decoupled approach, since the loss is taken over the posterior estimate for the latent parameter $\{\beta_t\}$, the MCMC samples can be utilized in identifying the optimal set of changepoints. 

First, minimization of the loss in \eqref{dec_loss} with our recommended penalty function can be solved via coordinate descent to produce a path of $\lambda$ values corresponding to different number of changepoints. This path of solutions will express a direct trade-off between goodness-of-fit and the number of changepoints. As the number of changepoints increases, the estimated solution $\widetilde{\pmb{\beta}}$ will be closer to the posterior mean across time. 

Second, for each $\lambda$ in the corresponding solution path, we will compute the ``projected posterior" \citep{woody2021model} to quantify its uncertainty. The key idea behind the projected posterior is to project each MCMC draw from the Bayesian model onto the summary space defined by locations of changepoints. For a given value $\lambda$, let $\eta$ denote the time indices which $\{\bigtriangleup^D \widetilde{\beta}_t \neq 0\}$ (i.e. the estimated changepoint locations). Initial points $1,\ldots,D$ are automatically included in every $\eta$ as they are unpenalized. Let $\pmb{\beta}^{(i)}$ denote the $i$th MCMC draw from the Bayesian model and $\pmb{Z}$ denote the inverse of the $Dth$ difference matrix. Let $\pmb{Z}_{\eta}$ denote the subset of columns of $\pmb{Z}$ indexed by a given $\eta$. The $i$th projected posterior is then given by:
\begin{equation}\label{eq:proj}
 \pmb{\beta}^{(i)}_\eta = (\pmb{Z}^{\mathbb{T}}_{\eta} \pmb{Z}_{ \eta})^{-1} \pmb{Z}^{\mathbb{T}}_{\eta} \pmb{\beta}^{(i)}  ,
\end{equation}
where $\mathbb{T}$ is the transpose operator. This projects $\pmb{\beta}^{(i)}$ from each MCMC draw onto the best fitted model given the changepoint estimates. In summary, the ``projected posterior" takes a set of changepoint locations and produces the best estimate of $\pmb{\beta}$ for each of the MCMC draws given the changepoint locations. This, in turn, allows us to visualize a trade-off between the number of changepoints and the corresponding fit for the posterior estimates.

Third, after deriving the projected posterior, we use a diagnostic tool to calculate a goodness-of-fit metric commonly based on amount of variation explained. Since we accounted for heteroskedasticity in the noise term of the Bayesian DLM, we propose using the following metric as an estimate for the amount of variation explained by the changepoints for the $i$th MCMC draw:
\begin{equation*}
   R_\eta^{2, (i)} \equiv 1 - \frac{\sum_{t=1}^n w_t (x_t \beta_t^{(i)}-  x_t \beta_{\eta, t}^{(i)})^2}{\sum_{t=1}^n w_t (x_t \beta_t^{(i)}-  x_t \mu_{\pmb{\beta}^{(i)}})^2} 
\end{equation*}
where $\mu_{\pmb{\beta}^{(i)}}$ is the mean over $t$ of the $i$th MCMC draw, $\pmb{\beta}^{(i)}$. This metric is similar to R-squared in that it measures the amount of variation explained by the projected posterior $\pmb{\beta}_{\eta}$ for each of the MCMC draws. However, the error for each time-step is multiplied by the corresponding weight value, giving time-steps with higher variances lower weights. This makes sense as we expect more uncertainty in regions of high noise volatility. In turn, this metric provides an estimate of variation explained for each of the MCMC draws. The higher the value of $R_\eta^{2, (i)}$, the better the fit of the ``projected posterior" to the $i$th MCMC draw. For selecting the optimal value of $\lambda$, we will select the lowest number of changepoints which the upper 90$\%$ credible interval for $\tilde{E}[R_\eta^{2}]$ exceeds a certain threshold. We find this simple selection criterion to be quite effective in empirical settings and easy to visualize. Details on the threshold selection will be given in Section \ref{sec_sim_dec}. 

\begin{figure}[t!]
\setlength\extrarowheight{-5pt}
\setlength{\tabcolsep}{2pt}
  \centering
   \begin{tabular}{c c}
  Bayesian DLM & Projected Posteriors \\
  \includegraphics[scale=.1]{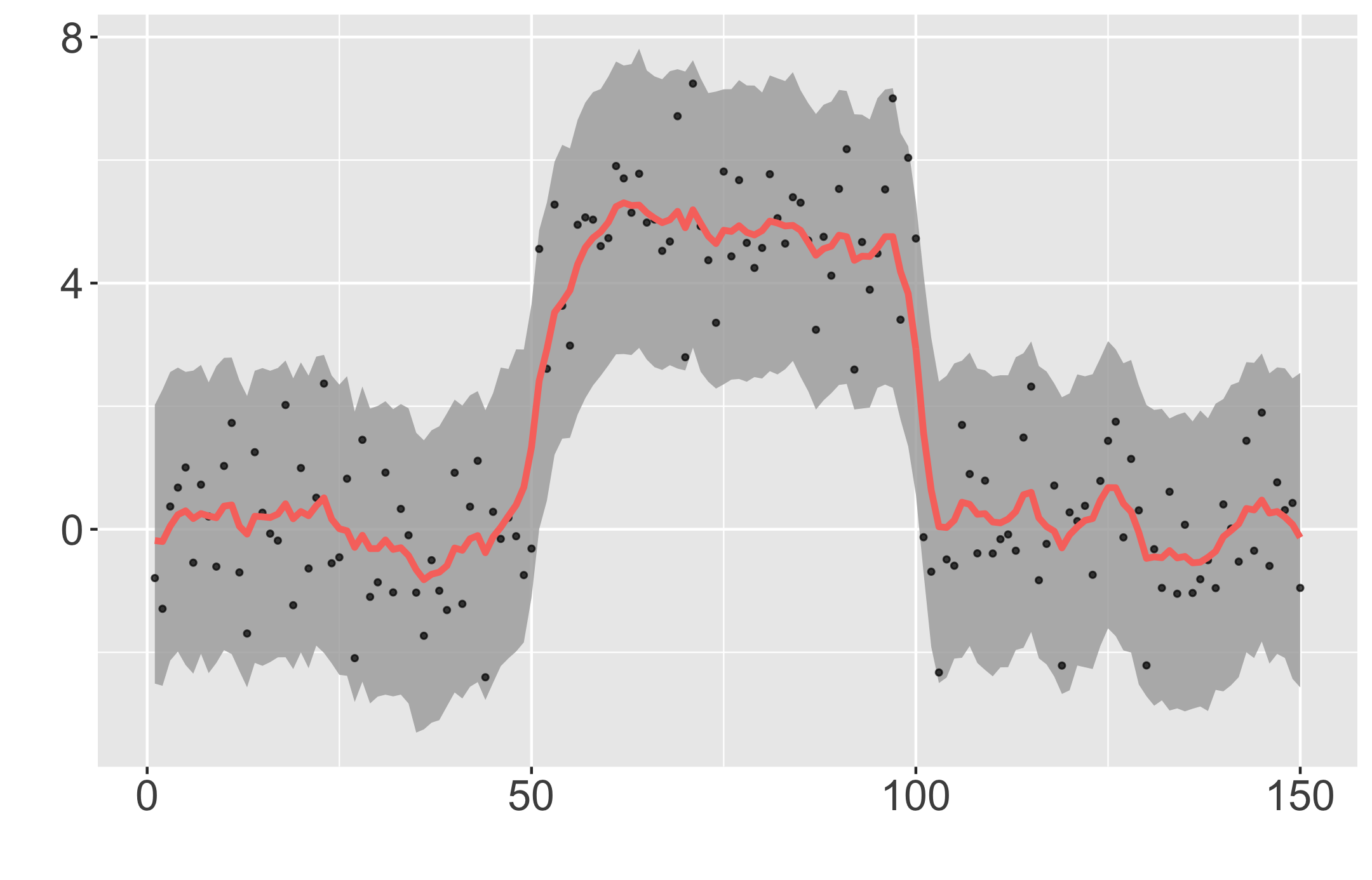} & \includegraphics[scale=.1]{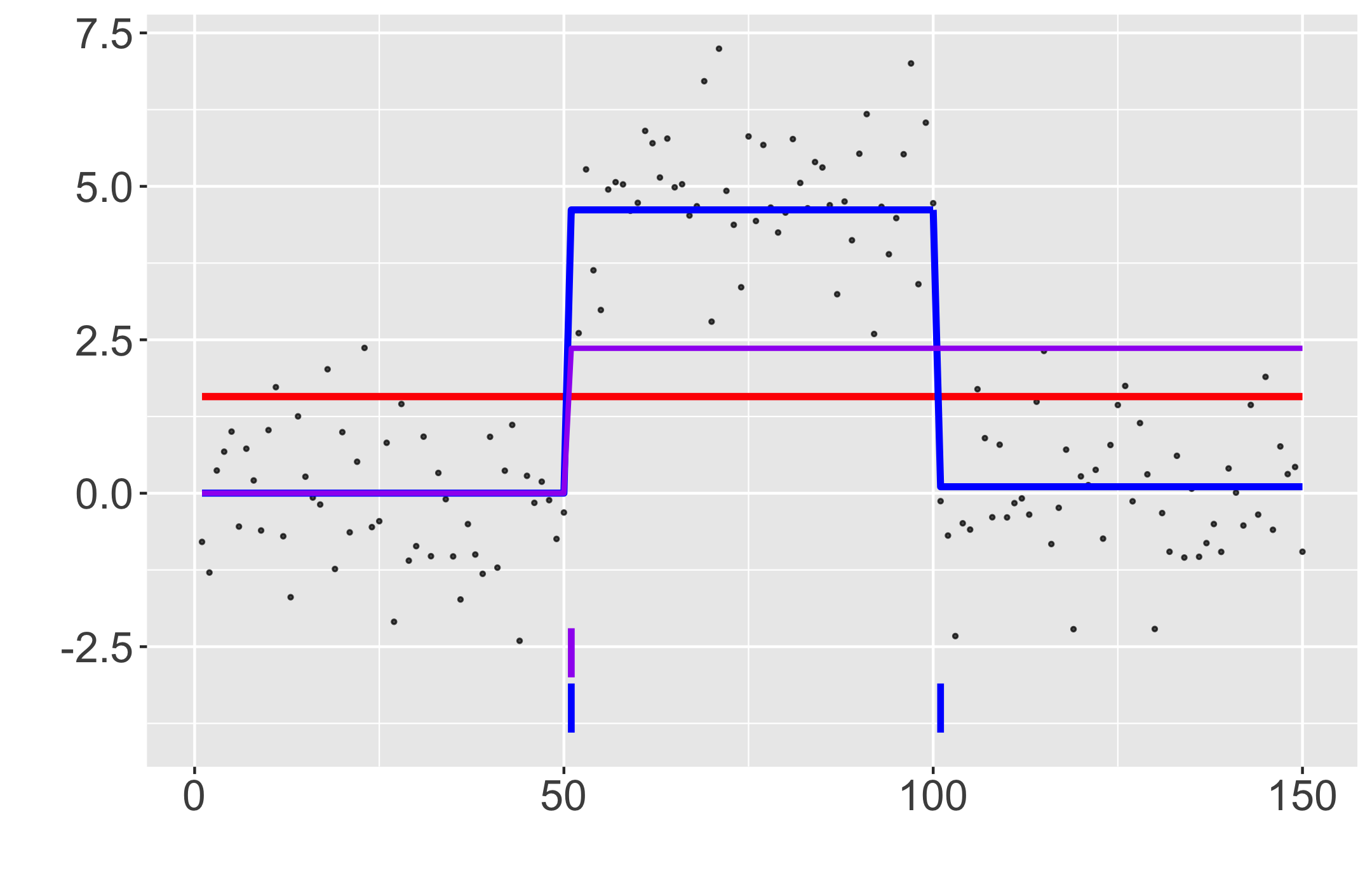} \\
  $R_\eta^2$ & Final Result \\
  \includegraphics[scale=.1]{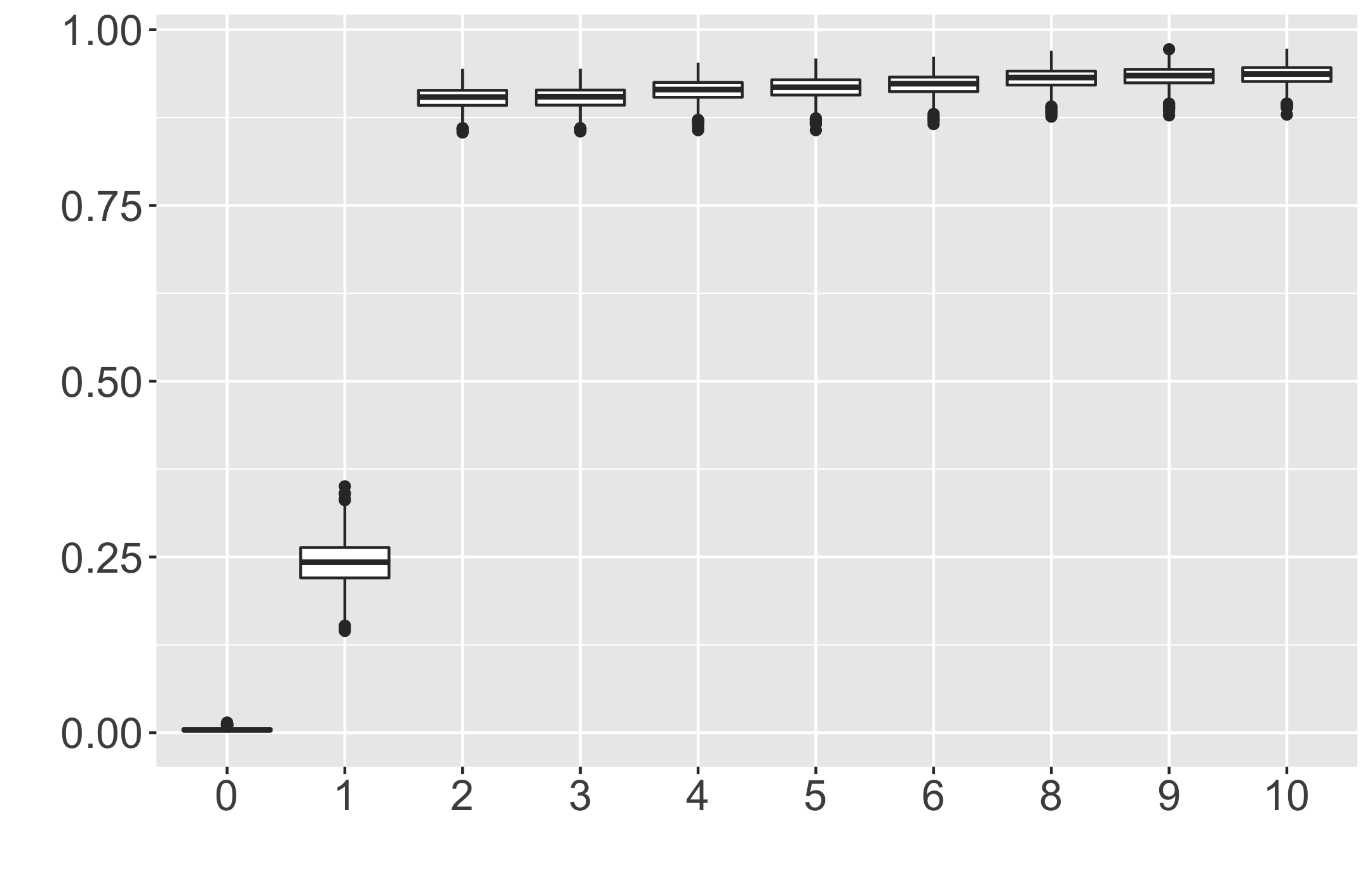} &
  \includegraphics[scale=.1]{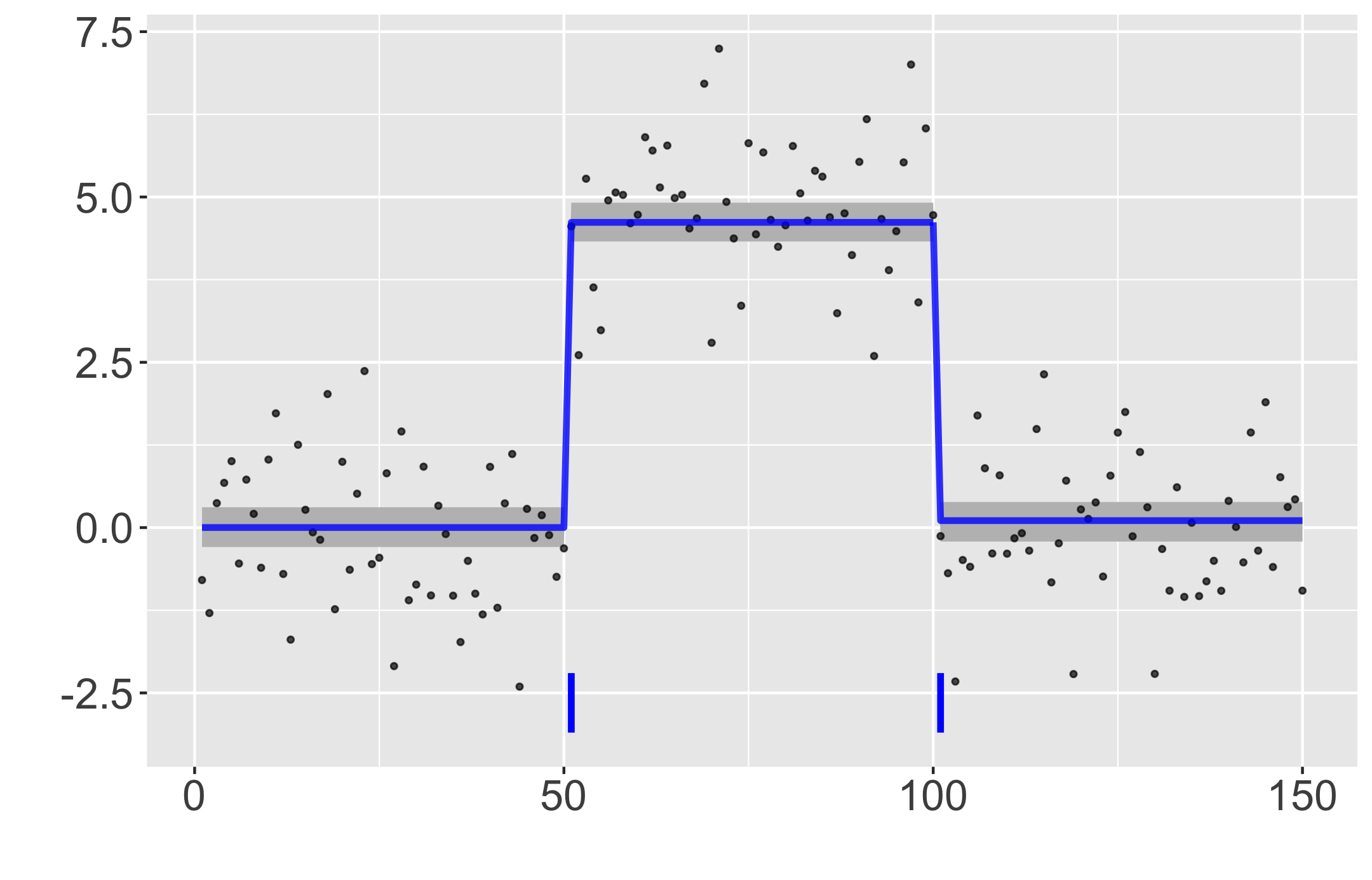} 
  \end{tabular}
  \caption{\setlength{\baselineskip}{1.0pt} Illustrative Example of the Decoupled Approach with Random Walk: The top-left plot shows the resulting posterior mean of $\{\beta_t\}$ using a Bayesian DLM with random walk. $90\%$ credible bands are shown as a ribbon. The top-right plot shows the mean of the projected posterior for 0, 1, and 2 number of changepoints. The bottom-left plots shows the distribution of $R_\eta^2$ as a function of the number of changepoints. The bottom-right plots shows the final resulting fit from the decoupled approach and 90\% bands corresponding to the projection.}
   \label{fig_rw_ex}
\end{figure}

Figure \ref{fig_rw_ex} illustrates the decoupled approach on a simulated series with two changepoints in mean. The fit from a Bayesian DLM with random walk is very wiggly but captures the underlying trend for the most part (Fig. \ref{fig_rw_ex} top-left). However, the model does not provide a clear identification of changepoints. The projected posterior for varying number of changepoints is shown in the top-right plot. For 0 changepoints, the projected posterior fits the global mean. For 1 changepoint, the projected posterior fits the first segment and combines the next 2 segments. For 2 changepoints, the projected posterior captures both true changepoints. This is reflected in the goodness-of-fit $R_\eta^2$. The metric shows large jumps from 0 to 1 changepoints and 1 to 2 changepoints, with marginal improvements afterward. We select 2 changepoints as the final result and plot the final projected posterior in the bottom-right plot. We additionally project all the posterior samples according to \eqref{eq:proj} for 2 changepoints (i.e., 3 non-zero values) and add the 90\% credible bands. The narrow uncertainty bands of the projection reflect the shrinkage of the penalization as compared to the uncertainty from the DLM. As seen, we can turn a very wiggly fit of the Bayesian DLM to a clear separation of drifts and shifts. 

\subsection{Locally Adaptive Trends with Global-Local Shrinkage Priors} \label{sec_shrinkage}
In the current model \eqref{eq1}, we assume the coefficients $\{\beta_t\}$ follow a random walk with a constant variance $\sigma_\omega^2$. While this setup can be sufficient for data with strong signal, this model tends to over-fit in datasets with low signal-to-noise ratios. Shrinkage priors present a trade-off between goodness-of-fit and smoothness of the underlying process; more shrinkage will typically result in a smoother underlying fit for the $\{\beta_t\}$ process. In this section, we will introduce the shrinkage priors for the decoupled approach. 

As previously discussed in Section \ref{sec_intro}, various forms of shrinkage priors have shown to be effective in Bayesian DLMs. For this section, we will focus on the class of so called ``global-local shrinkage priors" which have shown to be effective for Bayesian modeling \citep{bhadra2016default}. The prior on $\omega_{t}$ of equation \eqref{eq1} will be modified to $\omega_t \sim N(0, \tau_\omega^2 \gamma_{\omega, t}^2)$. This modification induces global-local shrinkage on the $D$th difference of the coefficients for the predictor. The parameter $\tau_\omega^2$ induces global shrinkage across all time-steps and the process $\{\gamma_{\omega, t}^2\}$ induces time-specific shrinkage for the coefficients. The two parameters combined shrink small deviations toward zero while allowing large signals to remain unchanged. This provides localized adaptivity while maintaining strong global shrinkage. 

In time dependent data, an additional dependence in the latent shrinkage or selection process has been shown to improve estimation in a variety of techniques \citep{nakajima2013bayesian, kowal_2018, Wu_2020, rockova2021dynamic}. One example is the dynamic shrinkage process detailed in \citet{kowal_2018}. The shrinkage process is detailed as follows:
\begin{equation*}\label{state}
h_t \equiv \log(\tau_\omega^2\gamma_{\omega, t}^2), \quad h_{t} = u + \phi(h_{t-1} - u) + \xi_{t},  
\end{equation*}
where $\phi$ is a univariate autocorrelation parameter, $\xi_{t} \stackrel{iid}{\sim} Z(0.5, 0.5, 0, 1)$, in which $Z(\cdot)$ denotes the four parameter $Z-$distribution, $Z(\alpha,\beta,\mu_z,\sigma_z)$, with density function
\begin{equation*}\label{eq:zdist}
    [z] = \{\sigma_zB(\alpha,\beta)\}^{-1}\exp\{(z-\mu_z)/\sigma_z\}^\alpha[1+\exp\{(z-\mu_z)/\sigma_z\}]^{-(\alpha+\beta)}, \; z \in \mathbb{R},
\end{equation*}
where $B(\cdot,\cdot)$ is the beta function. The distribution describes the log of an inverted beta random variable with parameters $\alpha$ and $\beta$. The parameters $\mu_z$ and $\sigma_z$ allow for shifting and scaling. Due to the previous effectiveness of the model, we utilize this model for all simulations with first differences (D = 1) unless otherwise specified and refer to it as decoupled dynamic shrinkage (DC-DS). 

Note that the decoupled approach is not restricted by the Bayesian DLM specification. A complex Bayesian model incorporating a variety of complexities such as covariates, heteroskedastic noise and non-stationary inputs can be fit to the data. The main recommendation is to fit a model that estimates fairly smooth coefficients for the predictors of interest. Then, the decoupled approach can adapt the inference part to identify key changepoints. This level of flexibility grants the decoupled approach the ability to work with more applications than previous existing changepoint algorithms. Extensions for dealing with multiple predictors and static parameters are shown in the Appendix 4.

\section{Simulated Experiments} \label{sec_sim_dec}
In this section, we illustrate the effectiveness and flexibility of the decoupled approach. The competing methods are the Pruned Exact Linear Time method \citep[PELT,][]{Killick_2012} and Robust FPOP algorithm \citep[R-FPOP,][]{Fearnhead_2019}. PELT identifies changepoint based on penalized cost function using a goodness-of-fit metric based on maximum negative likelihood for each segment and a penalty parameter on the number of changepoints. R-FPOP adapts the PELT penalty function using a biweight-loss in order to deal with outliers by establishing a maximum threshold for the impact of each time-step. The first simulation setting does not include an outlier process so therefore the R-FPOP method will not be considered a comparative method.

These two methods are similar to the decoupled approach in their utilization of a penalized cost function. However, unlike the decoupled approach, they utilize the data rather than posterior of a Bayesian model. The comparisons will start on simple cases of changes in mean with Gaussian noise, then extend to more complicated scenarios adding in outliers and heterogeneity. For both competing methods, we will use the default parameters as utilized in the original papers. For the decoupled approach, we use a cutoff threshold for 0.9 for lowest number of changepoints which the upper $90\%$ credible interval for $R_\eta^2$ exceeds. Full details of the parameters used for the Bayesian DLM and comparisons for other simulation settings are shown in the Appendix.

\subsection{Comparison Metric Details}

Five metrics are used to evaluate the results for simulations: Rand index, adjusted Rand index, precision, recall and F1-score. Rand index calculates a similarity score between the predicted partition and the true partition; the score ranges between 0 and 1 with 1 being a perfect match \citep{hubert1985comparing}. Adjusted Rand index provides an additional correction step to the Rand Index by accounting for random chance of a correct partition. Precision measures proportion of true changepoints in the number of predicted changepoints while recall measures proportion of all true changepoints detected by the models. F1-score calculates the harmonic mean between precision and recall. Since changepoints occur very rarely in the data, the F1-score is a good indicator for the accuracy of predictions \citep{van2020evaluation}. We consider a predicted changepoint to be a true positive if it is within $\pm 5$ of a true changepoint, with the caveat that each true changepoint can only match to at most one predicted changepoint. 

\subsection{Change in Mean with Gaussian Noise} \label{change_mean_gaussian}
For the first set of simulations, we start with a simple change in mean with standard Gaussian noise. We simulate data of length 200, with a changepoint in the middle of the data at location 100. We adjust different levels for the magnitude of change (MC) to understand the effectiveness of the algorithms with varying signal-to-noise ratios. We test 4 different magnitudes of change values of $\{1, 0.75, 0.5, 0.25\}$. We will compare the decoupled approach with a random walk state equation (DC-RW, model introduced in Section \ref{sec_rw}) and the decoupled approach with dynamic shrinkage (DC-DS, model introduced in Section \ref{sec_shrinkage}) against PELT. PELT is a penalized likelihood changepoint algorithm which is designed to identify changes in this setting, making it a good baseline for comparison. We expect the decoupled approach to perform slightly worse than PELT in this setting as a trade-off for increased flexibility. As we will show in the later simulations, the flexibility of the decoupled approach allows it to perform much better when the assumptions of homoskedasticity and Gaussian noise are violated.

\begin{table}[t!]
\caption{Single Change in Mean}
\label{tab_mean_change_gaussian}
\setlength{\baselineskip}{1.0pt} 
\setlength\extrarowheight{-11pt}
\centering
\begin{tabular}{| c c c c c c c|}
\hline 
  MC & Algorithms & Rand Avg. & Adj. Rand Avg. & Precision & Recall & F1-score \\ 
 \hline
  1 & DC-RW & $0.795_{(0.013)}$ & $0.590_{(0.027)}$ & 0.18 & 0.83 & 0.29 \\
  & DC-DS & $0.964_{(0.004)}$ & $0.929_{(0.009)}$ & 0.78 & 0.83 & 0.80 \\
  & PELT & $\pmb{0.968}_{(0.004)}$ & $\pmb{0.936}_{(0.007)}$ & \pmb{0.82} & \pmb{0.87} & \pmb{0.84} \\
 \hline 
  0.75 & DC-RW & $0.689_{(0.016)}$ & $0.379_{(0.031)}$ & 0.14 & 0.53 & 0.22 \\
  & DC-DS & $0.926_{(0.009)}$ & $0.851_{(0.018)}$ & \pmb{0.68} & 0.56 & 0.61 \\
  & PELT & $\pmb{0.930}_{(0.009)}$ & $\pmb{0.860}_{(0.021)}$ & 0.62 & \pmb{0.67} & \pmb{0.64} \\
  \hline
  0.5 & DC-RW & $0.576_{(0.012)}$ & $0.156_{(0.024)}$ & 0.12 & 0.25 & 0.16 \\
  & DC-DS & $\pmb{0.791}_{(0.017)}$ & $\pmb{0.582}_{(0.035)}$ & 0.23 & \pmb{0.38} & 0.29 \\
  & PELT & $0.755_{(0.021)}$ & $0.512_{(0.041)}$ & \pmb{0.38} & 0.30 & \pmb{0.33} \\
  \hline
  0.25 & DC-RW & $0.498_{(0.000)}$ & $0.000_{(0.000)}$ & 0.00 & 0.00 & 0.00 \\
  & DC-DS & $0.510_{(0.006)}$ & $0.025_{(0.013)}$ & 0.02 & 0.01 & 0.01 \\
  & PELT & $0.498_{(0.000)}$ & $0.000_{(0.000)}$ & 0.00 & 0.00 & 0.00 \\
  \hline 
\end{tabular}
\begin{flushleft}
  \setlength{\baselineskip}{1.0pt} 
  {Table details results of the decoupled approach with random walk (DC-RW), decoupled approach with dynamic shrinkage (DC-DS), and PELT on simulated data with one change in mean of varying magnitudes (MC) and standard Gaussian noise. Rand average and adjusted Rand average measures the similarity between predicted partition and true partition. Standard error for Rand average and adjusted Rand average are given in subscripts. F1-score measures accuracy of changepoint detection through a comparison of precision and recall. Bolded values indicate best results for the metric in the column.}
\end{flushleft}
\end{table}

As seen in Table \ref{tab_mean_change_gaussian}, the decoupled approach with dynamic shrinkage performs slightly worse than PELT. With a signal-to-noise ratio of 1 to 1 (magnitude of change 1), both DC-DS and PELT perform similarly well with Rand average above 0.95, adjusted Rand average above 0.925 and F1-score above 0.8. As the signal-to-noise ratio reaches a low of 1 to 4 (magnitude of change 0.25), both changepoint algorithms can no longer distinguish the correct changepoint. For the magnitude change of 0.75 and 0.5, PELT performs slightly better in terms of F1-score. However, we still see a trade-off of precision and recall between the two algorithms. For magnitude of change of 0.5, PELT has a higher precision while DC-DS has a higher recall. This shows that DC-DS has a tendency to slightly over-predict in low signal-to-noise ratio while PELT has a tendency to under-predict. A key note is that DC-DS maintains the highest Rand and adjusted Rand average in this settings, showing that DC-DS produces the partition closest to the true partition. 

Comparing DC-RW against DC-DS, we can clearly see that using shrinkage priors in the Bayesian DLM significantly improves the performance of the decoupled approach. This is due to the fact that shrinkage priors induce smoother estimates of the underlying trend resulting in easier changepoint inference. This further supports the discussion in Section \ref{sec_shrinkage} of the advantages of the decoupled framework allowing for fitting of any appropriately complex Bayesian model. Due to the significant improvements of using shrinkage priors in the baseline case, we will use DC-DS as our main method from this point onward.

\subsection{Change in Mean with Outliers} \label{change_mean_out}
For the next set of simulations, we added outliers onto the same problem as Section \ref{change_mean_gaussian} to illustrate the robustness of the methods. All other simulation settings will be kept the same as Section \ref{change_mean_gaussian}. As this is a more difficult problem, we increase the magnitude of changes to $\{2, 1.5, 1, 0.5\}$. Instead of Gaussian noise, we will utilize t-distributed noise with 2 degrees of freedom to simulate data with outliers. 

\begin{table}[t!]
\caption{Change in Mean with Outliers}
\label{tab_mean_change_outliers}
\setlength{\baselineskip}{1.0pt} 
\setlength\extrarowheight{-11pt}
\centering
\begin{tabular}{ |c c c c c c c|}
\hline 
  MC & Algorithms & Rand Avg. & Adj. Rand Avg. & Precision & Recall & F1-score \\ 
  \hline
  2 & DC-DS & $\pmb{0.977}_{(0.003)}$ & $\pmb{0.954}_{(0.007)}$ & \pmb{0.88} & \pmb{0.91} & \pmb{0.89} \\
  & PELT & $0.681_{(0.006)}$ & $0.361_{(0.012)}$ & 0.06 & 0.82 & 0.11 \\
  & R-FPOP & $0.952_{(0.011)}$ & $0.904_{(0.022)}$ & 0.86 & 0.83 & 0.84 \\
  \hline 
  1.5 & DC-DS & $\pmb{0.967}_{(0.005)}$ & $\pmb{0.934}_{(0.009)}$ & \pmb{0.80} & \pmb{0.82} & \pmb{0.81} \\
  & PELT & $0.689_{(0.007)}$ & $0.376_{(0.014)}$ & 0.05 & 0.74 & 0.10 \\
  & R-FPOP & $0.860_{(0.020)}$ & $0.721_{(0.040)}$ & \pmb{0.80} & 0.63 & 0.70  \\
  \hline
  1 & DC-DS & $\pmb{0.935}_{(0.007)}$ & $\pmb{0.870}_{(0.015)}$ & \pmb{0.68} & 0.60 & \pmb{0.63} \\
  & PELT & $0.680_{(0.007)}$ & $0.358_{(0.013)}$ & 0.04 & 0.59 & 0.08 \\
  & R-FPOP & $0.804_{(0.009)}$ & $0.638_{(0.019)}$ & 0.35 & \pmb{0.62} & 0.44 \\
  \hline
  0.5 & DC-DS & $\pmb{0.742}_{(0.018)}$ & $\pmb{0.486}_{(0.036)}$ & \pmb{0.18} & 0.32 & \pmb{0.24} \\
  & PELT & $0.676_{(0.006)}$ & $0.350_{(0.013)}$ & 0.04 & \pmb{0.48} & 0.07 \\
  & R-FPOP & $0.741_{(0.017)}$ & $0.484_{(0.035)}$ & 0.15 & 0.25 & 0.19 \\
 \hline 
\end{tabular}
\begin{flushleft}
  \setlength{\baselineskip}{1.0pt} 
  {Table details decoupled approach with dynamic shrinkage (DC-DS), PELT, and R-FPOP on simulated data with one change in mean of varying magnitudes (MC) and outliers. Outliers are simulated using t-distributed noise with 2 degrees of freedom. Rand average and adjusted Rand average measures the similarity between the predicted partition and true partition. Standard error for Rand average and adjusted Rand average across simulations are given in subscripts. F1-score measures accuracy of changepoint detection through a comparison of precision and recall. Bolded values indicate best results for the metric in the column.}
\end{flushleft}
\end{table}

The results of the simulation can be seen in Table \ref{tab_mean_change_outliers}. With the addition of outliers, the decoupled approach is able to achieve the best performance across all settings. By utilizing a Bayesian DLM with dynamic shrinkage, the decoupled approach is robust to the presence of extreme outliers. Unsurprisingly, PELT, with no mechanism to deal with extreme values, is significantly influenced by outliers. This lead to PELT significantly over-predicting the number of changepoints. In the setting of magnitude of change of 2, DC-DS achieves an F1-score of 0.89 in comparison to 0.84 for R-FPOP. As the signal-to-noise ratio decreases from 2 to 1, we can see an increasing gap in adjusted Rand average and F1-score between DC-DS and R-FPOP. This indicates that DC-DS can produce more precise changepoint estimations and more accurate partitions. The advantage becomes more significant in the setting of magnitude of change of 1. DC-DS achieves an F1-score of 0.63 in comparison to 0.44 of R-FPOP. As magnitude of change approaches 0.5, the problem becomes too difficult for all algorithms and performance is comparable between DC-DS and R-FPOP. 

\begin{figure}[t!]
\setlength{\baselineskip}{1.0pt}
\centering
    \begin{tabular}{c c c}
        Gaussian Noise & Outliers & Stochastic Volatility \\
        \includegraphics[scale = 0.09]{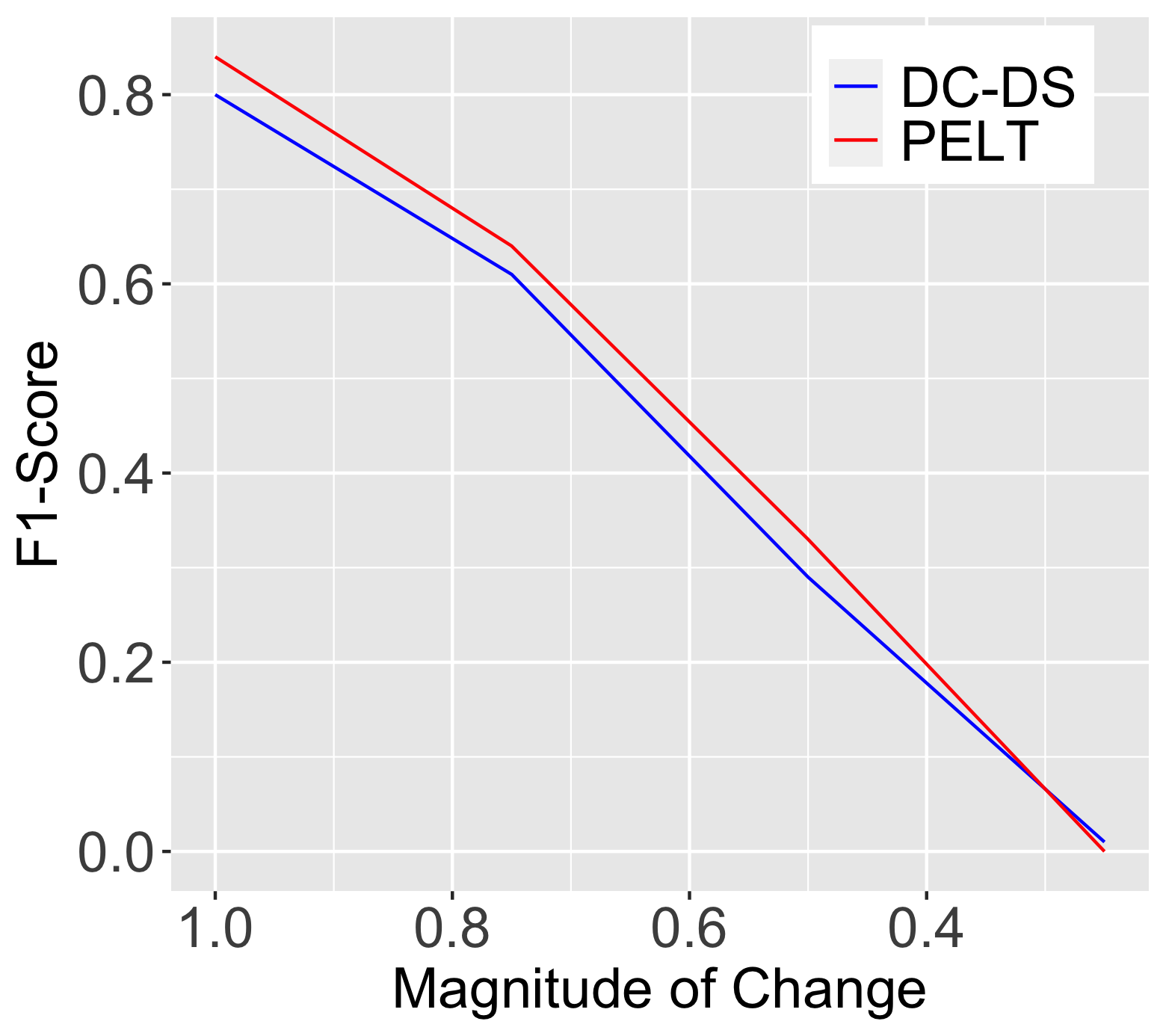} & \includegraphics[scale = 0.09]{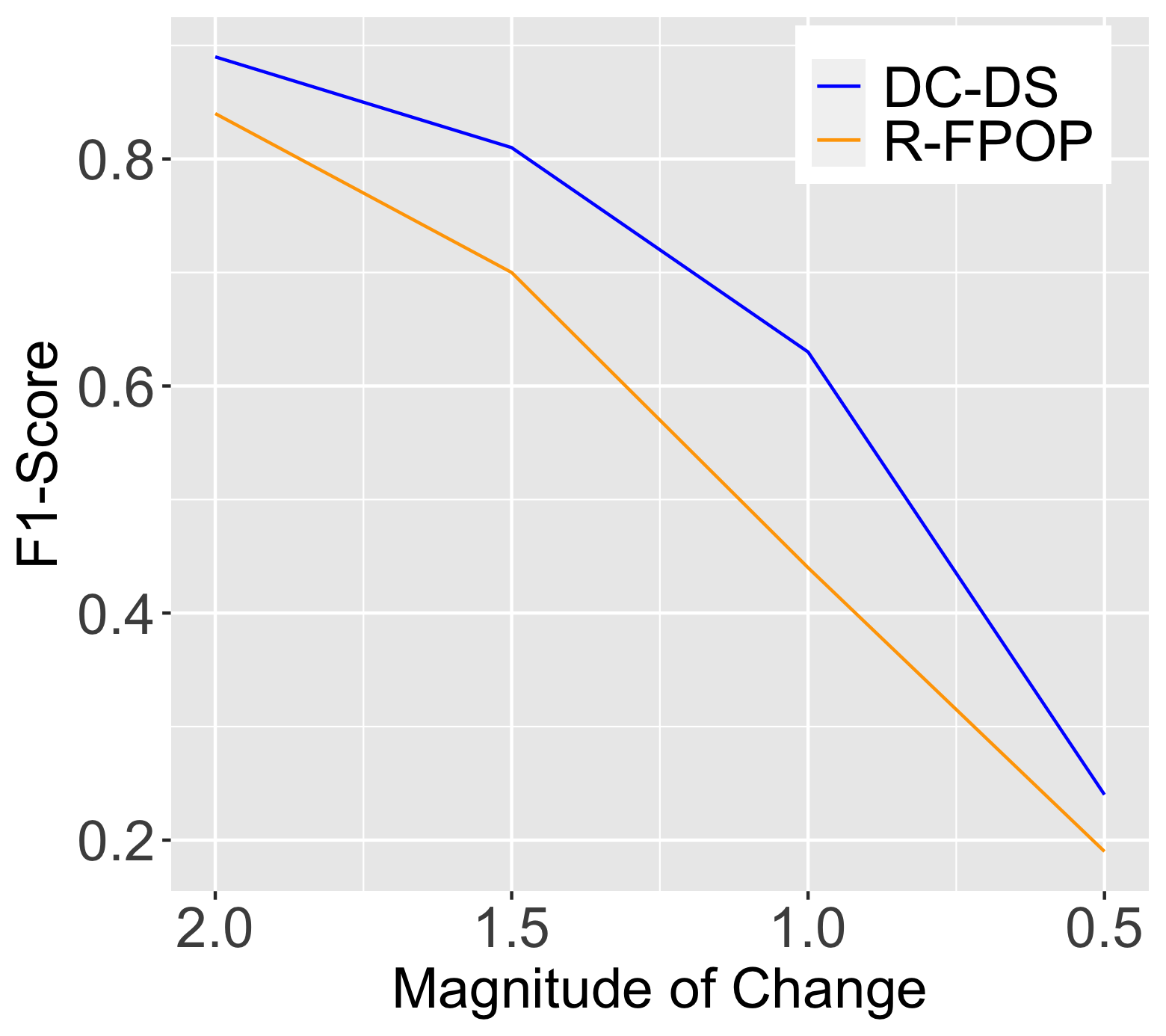} &
        \includegraphics[scale = 0.09]{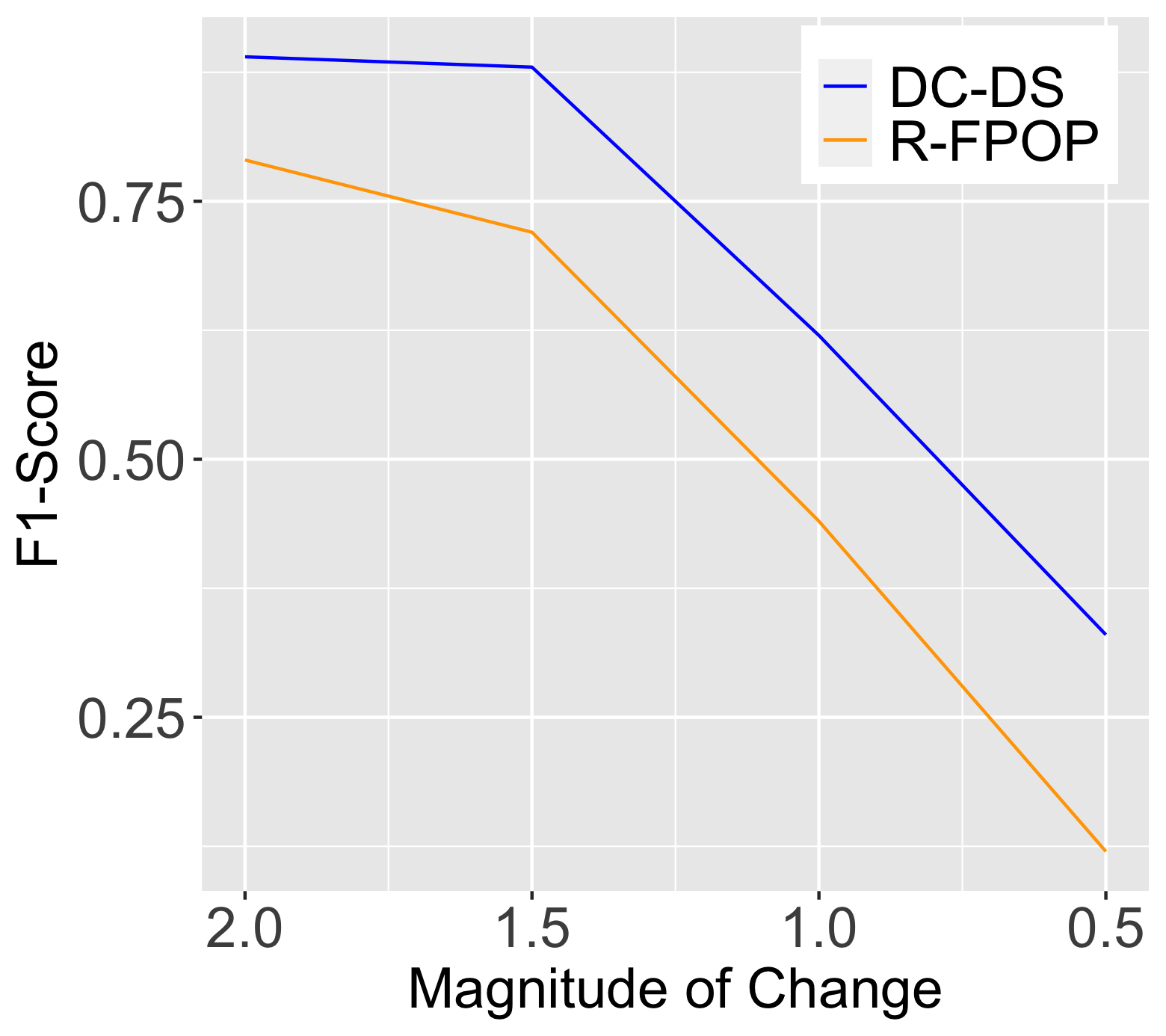} \\
    \end{tabular}
    \begin{flushleft}
    \vspace{-24pt}
   \caption{\setlength{\baselineskip}{1.0pt}  Change in Mean, comparison of F1-Scores. F1-score calculates the harmonic mean between precision and recall. The score ranges between 0 and 1 with 1 being a perfect prediction. The left plot shows F1-score of DC-DS against PELT in simulated data with Gaussian noise from Section \ref{change_mean_gaussian}. The middle plot shows F1-score of DC-DS against R-FPOP in simulated data with outliers from Section \ref{change_mean_out}. The right plot shows F1-score of DC-DS against R-FPOP in simulated data with stochastic volatility from Section \ref{change_mean_sv}.}
   \end{flushleft}
   \label{fig_f1_comp_mean}
\end{figure}

\subsection{Change in Mean in Presence of Heteroskedasticity} \label{change_mean_sv}
For the next set of simulations, we evaluate these algorithms in presence of heterogeneity. We simulate 100 series of length 200, with a changepoint in the middle of the data at location 100. However, instead of standard Gaussian noise or t-distributed noise, we generate noise using stochastic volatility of order 1 \citep{kim_1998} as follows:
\begin{equation} \label{eq_sv}
\log(\sigma_{\epsilon, t}^2) = \mu_\epsilon + \phi_\epsilon[ \log(\sigma_{\epsilon, t-1}^2)-\mu_\epsilon]+\xi_{\epsilon, t}, \; \; \; \; \; \; \; \; \; \; \; \; \xi_{\epsilon, t} \sim N(0, \sigma_{\eta}^2).
\end{equation}
We set the following values: $\mu_\epsilon = 0$, $\phi_\epsilon = 0.9$, and $\sigma_{\epsilon, t}^2 = 0.5$. This creates high auto-correlation which causes regions of high/low volatility which can occur frequently in real world data. We utilized 4 magnitude of change values of $\{0.5, 1, 1.5, 2\}$ to evaluate the algorithms' effectiveness in varying signal-to-noise ratios. The results are reported in Table \ref{tab_mean_change_hetero}. To be fair to PELT and R-FPOP, the algorithms are not intended to work in this setting. As a result, the performances are not reflective of the effectiveness of the algorithms. 

\begin{table}[t!]
\caption{Change in Mean with Heterogeneity}
\label{tab_mean_change_hetero}
\setlength{\baselineskip}{1.0pt} 
\setlength\extrarowheight{-11pt}
\centering
\begin{tabular}{ |c c c c c c c|}
\hline \hline
  MC & Algorithms & Rand Avg. & Adj. Rand Avg. & Precision & Recall & F1-score \\ 
  \hline
  2 & DC-DS & $\pmb{0.982}_{(0.004)}$ & $\pmb{0.964}_{(0.008)}$ & \pmb{0.89} & 0.90 & \pmb{0.89} \\
  & PELT & $0.834_{(0.010)}$ & $0.667_{(0.019)}$ & 0.11 & \pmb{0.96} & 0.20 \\
  & R-FPOP & $0.944_{(0.009)}$ & $0.889_{(0.018)}$ & 0.73 & 0.85 & 0.79 \\
  \hline
  1.5 & DC-DS & $\pmb{0.977}_{(0.004)}$ & $\pmb{0.955}_{(0.008)}$ & \pmb{0.87} & \pmb{0.90} & \pmb{0.88} \\
  & PELT & $0.827_{(0.010)}$ & $0.654_{(0.020)}$ & 0.10 & 0.89 & 0.19 \\
  & R-FPOP & $0.926_{(0.011)}$ & $0.852_{(0.022)}$ & 0.64 & 0.81 & 0.72 \\
  \hline 
  1 & DC-DS & $\pmb{0.931}_{(0.009)}$ & $\pmb{0.862}_{(0.020)}$ & \pmb{0.60} & \pmb{0.64} & \pmb{0.62} \\
  & PELT & $0.800_{(0.010)}$ & $0.599_{(0.020)}$ & 0.07 & 0.61 & 0.13 \\
  & R-FPOP & $0.835_{(0.019)}$ & $0.671_{(0.037)}$ & 0.39 & 0.50 & 0.44 \\
  \hline 
  0.5 & DC-DS & $\pmb{0.836}_{(0.015)}$ & $\pmb{0.673}_{(0.030)}$ & \pmb{0.29} & \pmb{0.38} & \pmb{0.33} \\
  & PELT & $0.686_{(0.013)}$ & $0.373_{(0.026)}$ & 0.03 & 0.25 & 0.05 \\
  & R-FPOP & $0.626_{(0.018)}$ & $0.254_{(0.036)}$ & 0.12 & 0.11 & 0.12 \\
 \hline 
\end{tabular}
\begin{flushleft}
  \setlength{\baselineskip}{1.0pt} 
  {Table details decoupled approach with dynamic shrinkage (DC-DS), PELT, and R-FPOP on simulated data with one change in mean of varying magnitudes (MC) and stochastic volatility. Stochastic volatility is simulated using highly autocorrelated SV(1) model. Rand average and adjusted Rand average measures the similarity between predicted partition and true partition. Standard error for Rand average and adjusted Rand average are given in subscripts. F1-score measures accuracy of changepoint detection through a comparison of precision and recall. Bolded values indicate best results for the metric in the column.}
\end{flushleft}
\end{table}

Comparing results in Table \ref{tab_mean_change_hetero} to Table \ref{tab_mean_change_gaussian} for magnitude of change 1, we see that the performance of all changepoint algorithms decreased in presence of stochastic volatility. This is to be expected as stochastic volatility makes detection of changepoints much more difficult. With the addition of stochastic volatility, DC-DS outperformed other competing changepoint methods in all settings. DC-DS achieves an F1-score of 0.89 for setting of magnitude of change of 2, an F1-score of 0.62 for setting of magnitude of change of 1 and an F1-score of 0.33 for setting of magnitude of change of 0.5. DC-DS achieves the most accurate partitions by having the highest adjusted Rand average and the best trade-offs of precision/recall. This illustrates the robustness of the decoupled approach in dealing with heterogeneity. 

Figure \ref{fig_f1_comp_mean} summarizes the results in term of F1-score for Section \ref{sec_sim_dec}. As seen in the plots, the decoupled approach is slightly worse in standard Gaussian noise settings but performs significantly better when outliers or stochastic volatility are added. This illustrates the trade-off of the decoupled approach. By fitting a Bayesian DLM to the data first, the decoupled approach can be more locally adaptive to the complexities inherent in time series data. Outliers and heterogeneity are just two examples of the challenges that the decoupled approach can deal with. As long as the posterior estimates for the $\{\beta_t\}$ process remains relatively smooth, the decoupled loss can identify correct changepoint locations in a variety of complex scenarios. 

\section{Global Land Surface Air Temperature Anomaly} \label{sec_real_w}

For an illustrative application we consider monthly global land surface air temperature anomaly with reference period 1951-1980 in 0.01 degrees Celsius from 1880 to 2018 (\url{https://data.giss.nasa.gov/gistemp/tabledata_v3/GLB.Ts.txt}). The urgency to detect sudden shifts in climate patterns has been growing amidst ongoing human-induced change. As shown, there are clear long term linear time trends underlying local annual trends in the data. Overall global temperatures appear increasing over time; however, three features make standard changepoint analysis difficult. First, there exists seasonal fluctuation in the data, and these seem somewhat irregular. This implies the local trend is not flat but rather a smooth curve fluctuating through the months. Second, there is differing levels of variability over time. Third, there may be anomalies  throughout the data as a result of certain global events. 

\begin{figure}[t!]
  \centering
  \setlength{\baselineskip}{1.0pt} 
  \setlength\extrarowheight{-9pt}
  \begin{tabular}{c c}
  Bayesian DLM Results & Projected Posteriors \\
  \includegraphics[width = 0.45\textwidth]{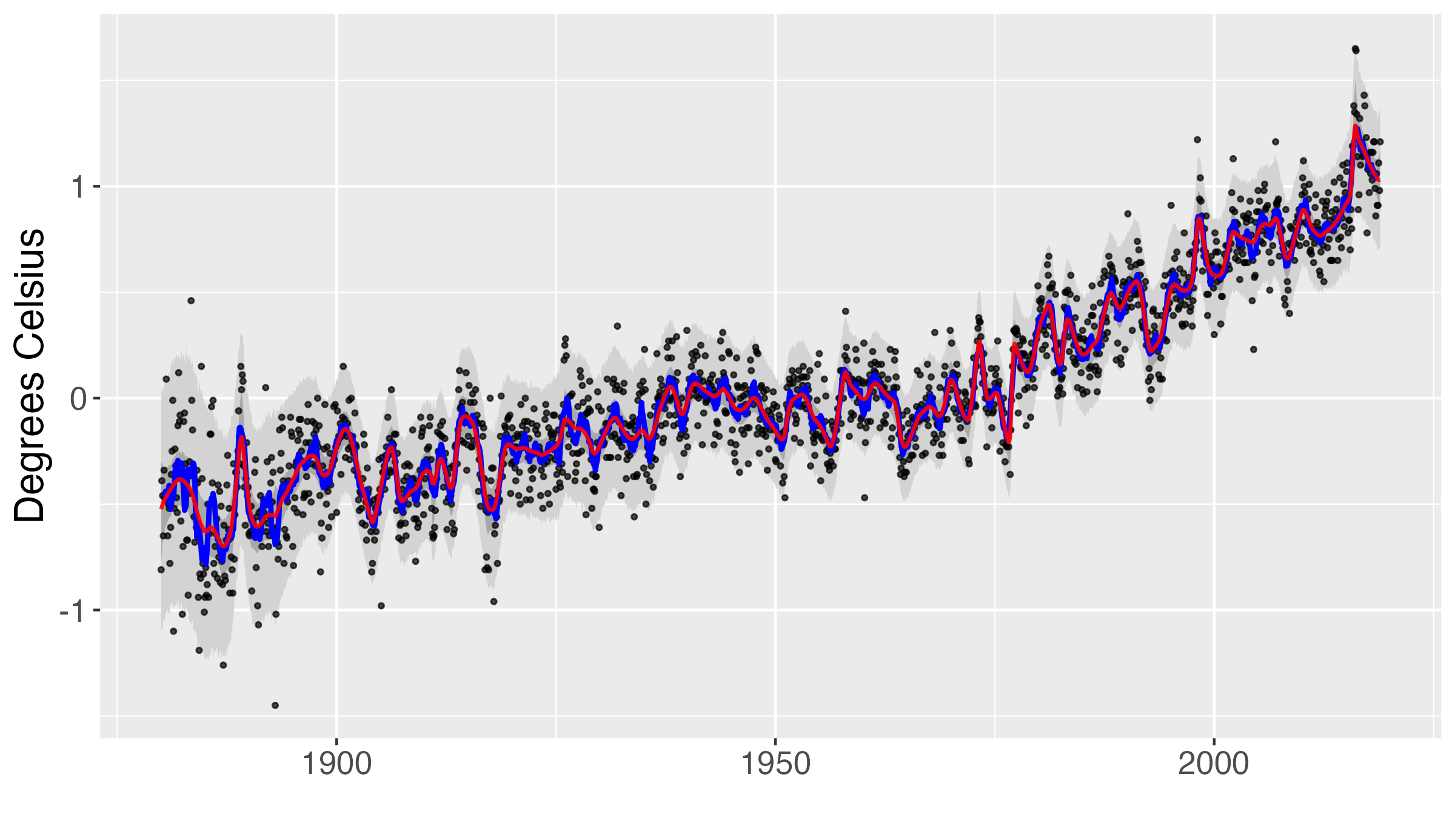} & \includegraphics[width = 0.45\textwidth]{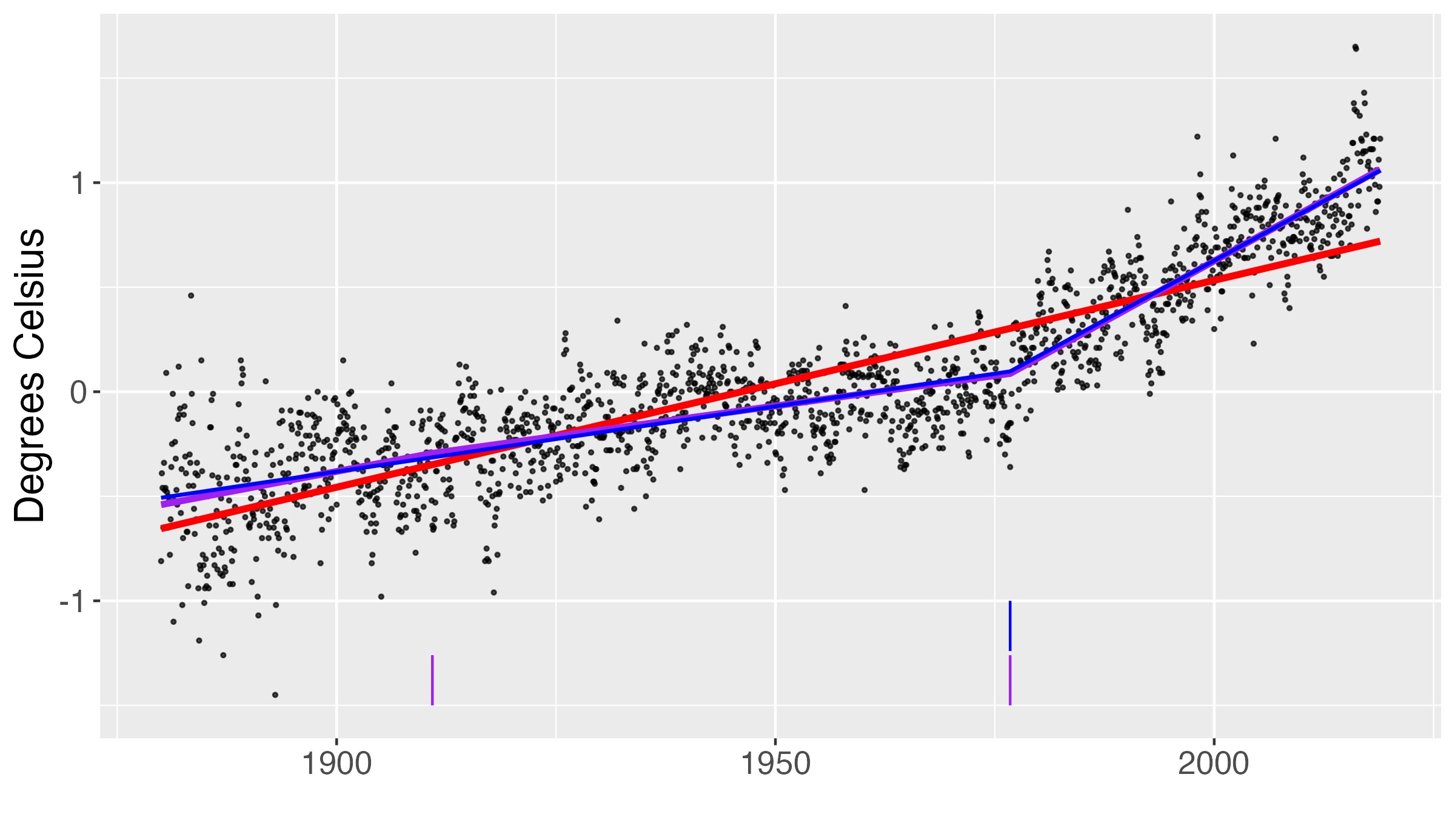} \\
  $R_\eta^{2}$ & Final Result \\
  \includegraphics[width = 0.45\textwidth]{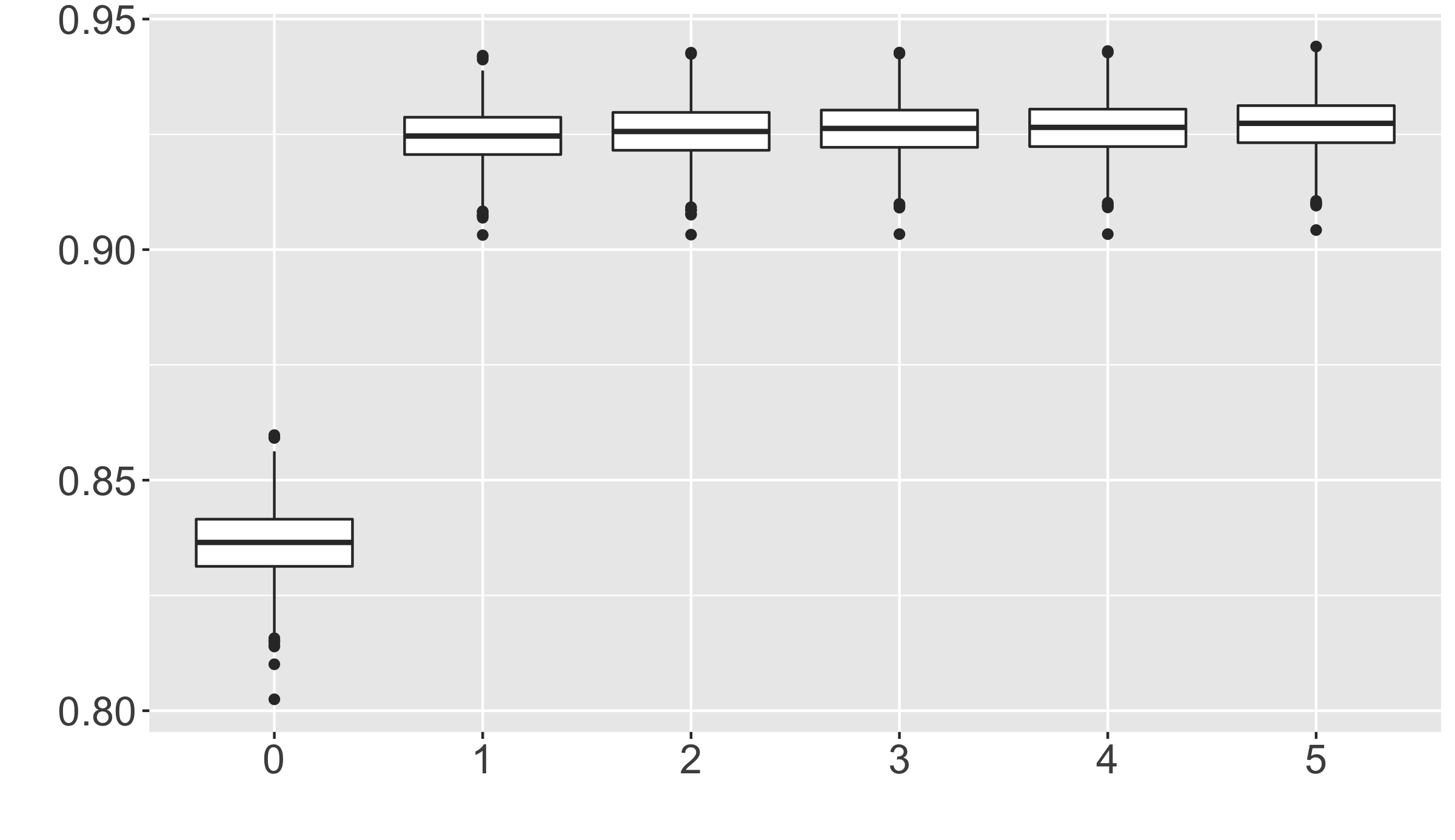} &
  \includegraphics[width = 0.45\textwidth]{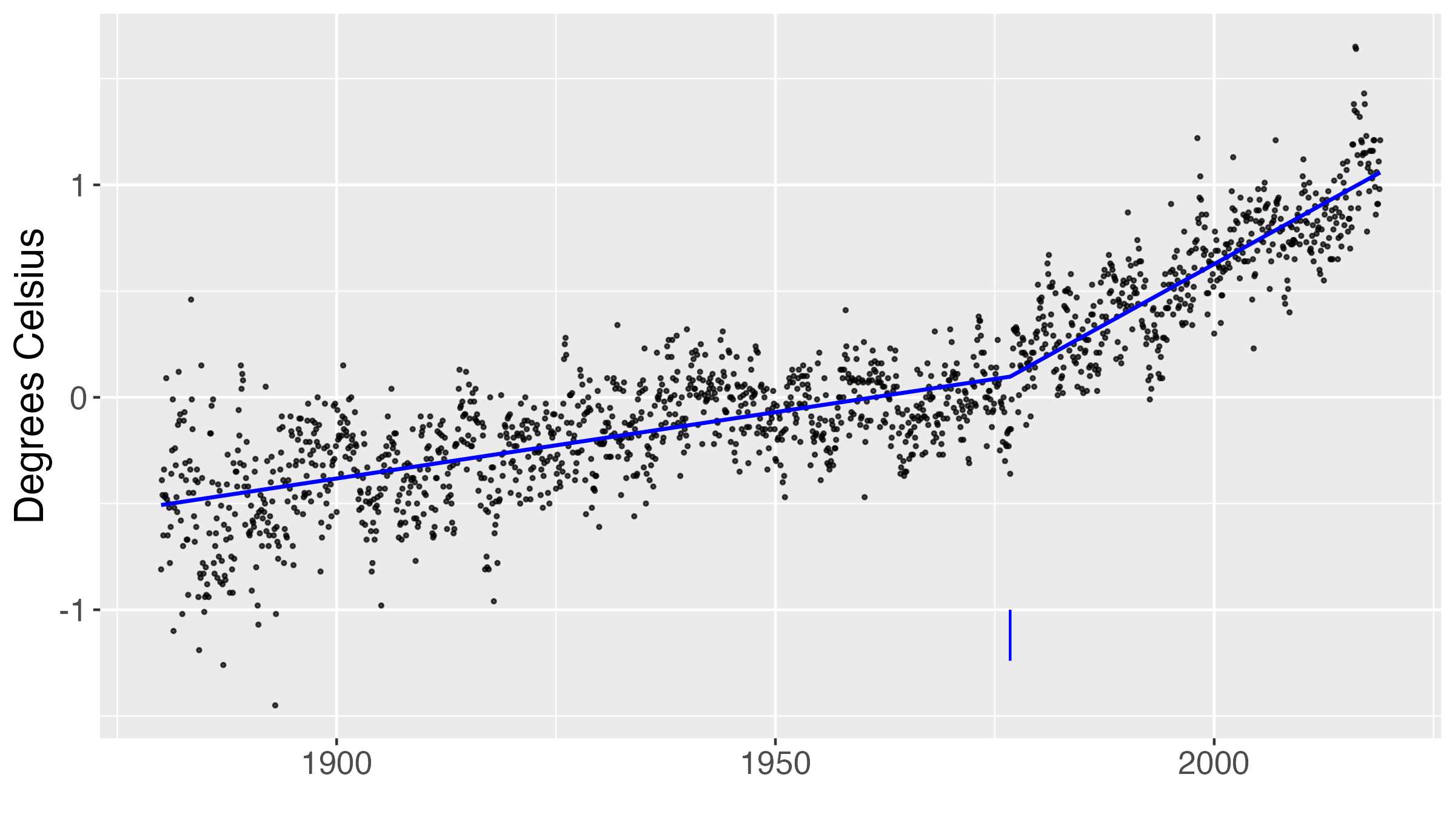} 
  \end{tabular}
  \begin{flushleft}
  \caption{\setlength{\baselineskip}{1.0pt} Monthly Global Land Surface Air Temperature Anomaly: The top-left figure shows the monthly average global land surface air temperature anomaly from 1880 to 2018 with 10 year moving average and Bayesian DLM fit. Additionally, the inner ribbon is 95$\%$ credible bands for $\{\beta_t\}$ and the outer ribbon is 95$\%$ credible bands for $\{\beta_t + \epsilon_t\}$ from the Bayesian DLM. The top-right figure shows the mean of ``projected posterior" for $\{0, 1, 2\}$ changepoints. The predicted changepoint locations are shown by the vertical lines. The bottom left plot illustrates the distribution $R_\eta^2$ for various number of changepoints. The bottom right plot shows the final result for the decoupled approach.}
   \end{flushleft}
   \label{figGLSAT}
\end{figure} 

As seen in the top-left plot of Figure \ref{figGLSAT}, the underlying signal fluctuates over time as a result of irregular cyclical patterns over the years; these patterns have less variability than the longer term approximately linear time trends. This results in a wiggly fit from the Bayesian dynamic linear model with $D=2$. Using the decoupled approach, we can visualize different fits of the projected posterior. The top-right plot of Figure \ref{figGLSAT} illustrates the mean of the ``projected posterior" for $\{0,1,2\}$ number of changepoints. As the number of changepoints increase, the fit becomes increasingly better. This is because increasing the number of changepoints essentially increases the degrees of freedom for the ``projected posterior". We select 1 as the optimal number of changepoints as it's the simplest fit in which the upper $90\%$ credible interval exceeds the 0.9 threshold. We estimate the single changepoint at November, 1976, after which there is a steeper long term slope. Our changepoint time aligns well with a recognized regime shift in the 1976-1977 winter originally determined by climate scientists in the 1990s based on multiple signals, but attributed to the North Pacific \citep{hare2000empirical}. Several changepoint analyses of temperature anomalies or multiple climate measures identify at least one shift in the 1970s decade \citep{alley2003abrupt, ivanov20101963, matyasovsky2011detecting, Yang_2014}. More real world applications involving changes in dynamic regression are shown in Appendix 5.

\section{Conclusion}
In conclusion, this paper proposes a decoupled approach for changepoint analysis that separates the processing of modeling and inference. As seen throughout the simulations and real world examples, the decoupled approach offers several key advantages over the competing method. First, by separating the process of modeling and inference, the decoupled approach allows for fitting of a highly complex Bayesian model to the underlying data while still allowing for reasonable inference of changepoints. This allows the decoupled approach to deal with many complexities inherent in time series. As the data becomes increasing complex, the decoupled approach can adapt the Bayesian DLM to deal with these issues while maintaining the same inference process for changepoints. Additionally, Bayesian modeling frameworks for other challenging time series data such as data with varying degrees of sparsity can be used in the first stage of the decoupled approach as long as it gives estimates of the trend at the desired inference times. 

Second, the decoupled approach is flexible in its ability to identify different types of changepoints. From the examples shown in the paper and the Appendix the decoupled approach has the ability to identify changes in mean, changes in regression coefficients and changes in higher order trends. Most other changepoint algorithms can only be utilized for one specific scenario. Lastly, the Bayesian decoupled approach maintains the ability to quantify uncertainty of parameters and derived quantities as compared to traditional changepoint algorithms. The flexibility of the decoupled approach allow the algorithm to be more adaptive to a wide variety of datasets and scientific conclusions. 

\bibliographystyle{plainnat}
\bibliography{paper}

\end{document}